\documentclass[fleqn,10pt]{wlscirep}
\usepackage{graphicx}
\usepackage{dcolumn}
\usepackage{bm}
\usepackage{amsmath}
\usepackage{caption}
\usepackage{subcaption}

\def\etal{\textit{et al.}} 

\DeclareRobustCommand{\_}[1]{
    \ifmmode
         \sb{\mathrm{#1}}
    \else
        \GenericError{\space\space\space\space}
            {Attempt to use \@backslashchar MyMathModeMacro outside of math mode}
            {See my preamble documentation for explanation.}
            {Need to use either use inline or display math.}%
    \fi
}

\title{The interfacial nature of proximity induced magnetism and the Dzyaloshinskii-Moriya interaction at the Pt/Co interface \\
}

\author[1,5]{R.M.~Rowan-Robinson}
\author[2]{A.A.~Stashkevich}
\author[2]{Y.~Roussign\'{e}}
\author[2]{M.~Belmeguenai}
\author[2]{S-M.~Cherif}
\author[3] {A.~Thiaville}
\author[4]{T.P.A.~Hase}
\author[1]{A.T.~Hindmarch}
\author[1,*]{D.~Atkinson}

\affil[1]{Department of Physics, University of Durham, Durham DH1 3LE, United Kingdom}
\affil[2]{Universite Paris 13, Sorbonne Paris Cite, 93430 Villetaneuse, France}
\affil[3] {Laboratoire de Physique des Solides, Université Paris-Sud, CNRS UMR 8502, 91405 Orsay Cedex, France}
\affil[4]{Department of Physics, University of Warwick, Coventry CV4 7AL, United Kingdom}

\affil[5]{Now at Department of Physics and Astronomy, Uppsala University, Box 516
751 20 Uppsala, Sweden}
\affil[*]{del.atkinson@durham.ac.uk}

\begin{abstract}
The Dzyaloshinskii-Moriya interaction (DMI) has been shown to stabilise Ne\'{e}l domain walls in magnetic thin films, allowing high domain wall velocities driven by spin current effects. DMI occurs at the interface between ferromagnetic and heavy metal layers with strong spin-orbit coupling, but details of the interaction remain to be understood and the role of proximity induced magnetism (PIM) in the heavy metal is unknown. We report interfacial DMI and PIM in Pt determined as a function of Au and Ir spacer layers in Pt/Co/Au,Ir/Pt. The length-scale for both interactions is sensitive to sub-nanometre changes in the spacer thickness, and they correlate over sub mono-layer spacer thicknesses, but not for thicker spacers. The spacer layer thickness dependence of the Pt PIM for both Au and Ir shows a rapid monotonic decay, while the DMI changes rapidly but has a two-step approach to saturation and continues to change, even after the PIM is lost.
\end{abstract}
\begin{document}

\flushbottom
\maketitle

\thispagestyle{empty}

Manipulating magnetisation with current is an extremely appealing prospect for magnetic memory and logic. First realized with spin-polarised current in spin-valves via  spin-transfer torque~\cite{Sankey2008}, progress has been limited by the high current densities required to induce domain wall motion~\cite{Parkin2008}. However, more recently the focus has moved towards spin-orbit torques, where the switching is driven by spin-current and leads to more efficient magnetisation control. Such magnetisation torques are most commonly attributed to the Rashba~\cite{Rashba1984,MihaiMiron2010} effect or spin-Hall effect~\cite{Hirsch1999,Ganguly2014} at the interface between the heavy metal and ferromagnetic layers.

Extremely high domain wall velocities have been observed in perpendicularly magnetized Pt/ferromagnetic films~\cite{Emori2012,Schellekens2012,Ryu2012,Haazen2013} and it has been suggested that these domain walls are driven by spin-current generated within the Pt and pumped across the interface into the ferromagnet via the spin-Hall effect. It has become clear that the primary spin-orbit torque associated with the spin-Hall effect is a damping-like torque with the same symmetry as the Slonczewski spin-transfer torque~\cite{Emori2013}. However the spin arrangement of the magnetostatically favoured Bloch walls in these systems is such that the damping-like torque is zero and the observed high domain wall velocities therefore require a N\'{e}el-type domain wall, which can be obtained in the presence of a Dzyaloshinskii-Moriya interaction (DMI)~\cite{Dzyaloshinskii1958,Moriya1960} across the interface. The DMI  stabilises a N\'{e}el type wall and imposes a chirality upon it~\cite{Thiaville2012}. The DMI exchange is given by $-\bm{D}\cdot \bm{S}\_{1} \times \bm{S}\_{2}$ and favours orthogonal alignment of spins $\bm{S}\_{1} $ and $\bm{S}\_{2}$ at the interface, which can be represented as an effective field acting across the wall that stabilises the Ne\'{e}l configuration. This effect also plays a crucial role in stabilising Skyrmion phases in magnetic thin films~\cite{Yu2010,Heinze2011,Boulle2016}.

In Pt/ferromagnet heterostructures, there also arises another phenomenon: a spontaneous magnetic polarisation in the interfacial region~\cite{Geissler2001}. This proximity induced magnetism (PIM) in Pt is associated with the large Stoner factor associated with higher \textit{d}-transition elements, with Pd, Ir and W all having been suggested to exhibit some degree of induced moment when placed in proximity to a ferromagnet~\cite{Clogston1962,Hase2014,Wilhelm2001,Stahler1993}.\\

To date, most experiments have focused on understanding the above spin-orbit phenomena~\cite{Jamali2013,Martinez2013}, and neglected the polarisation of the heavy metal layer. Attempts to distinguish between the Rashba and spin Hall components of spin-orbit torques have indicated an anomalous interfacial contribution~\cite{Fan2014,HinSim2014,Garello2013}, and also that the structure, as well as the spin-transparency, of the Pt/Co interface can dramatically modify the spin-orbit torque efficiency or the effective spin-Hall angle~\cite{Zhang2015,Tokac2015}. However, the specific role of PIM on the magnetisation dynamics and the relationship to DMI has been highlighted by recent work, where high domain wall velocities were only observed when heavy metal layers thought to exhibit PIM were used~\cite{Ryu2013}. Following this, ultrathin heavy metal spacer layers (SLs), notably Au, which exhibits an extremely weak proximity polarisation, were inserted between the Pt/ferromagnetic interface and displayed a linkage between PIM, DMI and domain mobility. These studies suggest there is an intimate link between DMI and PIM~\cite{Ryu2014}, however, in contrast, a recent theoretical work found no direct correlation between DMI and PIM in Pt~\cite{Yang2015}. To gain further physical insights into any possible relationship between DMI and PIM, and to address the conflict between previous experiments~\cite{Ryu2014} and theory~\cite{Yang2015}, requires direct measurements of both the DMI and PIM in the same samples. Furthermore, the analysis of the magnetic phenomena  needs to be considered in the context of the interfacial structure, the details of which are determined by the arrangment and interactions of the atomic consituents across the interface.\\

Here, we report an experimental analysis of both the PIM in Pt and the total DMI in Pt/Co/SL/Pt multilayers, where SL is a spacer layer of Au or Ir, and the analysis was performed as a function of SL thickness. The DMI and PIM measurements were also correlated using x-ray analysis of the physical structure of the multilayers and their interfaces. The elements for the spacer layer were selected based on previous understanding, where Au was selected following Ref.~\cite{Ryu2013,Ryu2014} as it is expected to have a negligible PIM~\cite{Wienke1991,burn2014focused}, and no DMI has been reported. In contrast, it has been suggested that Ir takes on a moderate PIM~\cite{Stahler1993} and a non-negligible interface DMI (IDMI) constant, as  was recently reported for a Co/Ir structure measured by Brillouin spectroscopy~\cite{Kim2016}. The difference in magnetic behaviour is not surprising considering the orbital filling for Ir and Au.\\

Multilayers of Pt(5.4)/Co(2.5)/SL(0-2.5)/Pt(2.6) (thicknesses in nm; with the convention that multilayers are described starting from the substrate) were sputter-deposited, with a spacer layer of either Au or Ir. The cobalt thickness was selected to produce in-plane magnetisation, which was confirmed by the magnetic hysteresis measured using MOKE.  In-plane magnetisation enables both PIM measurements and DMI analysis on the same samples.

\section*{Interfacial Dzyaloshinskii-Moriya interaction  as a function of Au and Ir spacer layers}

The DMI was measured by Brillouin light scattering (BLS)~\cite{Stashkevich2015}. BLS probes the dispersion relation of excitations, thermally populated here, through its momentum (wave number) and energy (frequency) resolution. The influence of DMI on the spin wave (SW) spectrum is now well-known, both theoretically~\cite{Udvardi2009,Cortes2013,Moon2013,Kost2014} and experimentally~\cite{Zakeri2010,Di2015,Belmeguenai2015,Cho2015}. DMI induces a characteristic non-reciprocity of the SW propagation, such that SWs with the same wave-number, but traveling in opposite directions have different frequencies. The symmetry of interfacial DMI implies that the SWs have to be probed in the so-called Damon-Eshbach (DE) geometry~\cite{Gurevich1996}, where the magnetisation and the wave-vector are in-plane and mutually perpendicular. In this configuration the frequency shift of the two counter-propagating modes is largest, and scales characteristically with the inverse of the sample thickness, which eliminates the contribution to the non-reciprocity due to the conventional asymmetric surface magnetic anisotropy~\cite{Stashkevich2015}. The two counter-propagating modes are recorded in the same spectrum, one as an energy loss (a frequency downshifted Stokes mode) and the other as an energy gain (a frequency upshifted anti-Stokes mode). Spectra are obtained by counting photons for typically 12 hours, so that the mode line positions can be determined with a precision better than 0.1~GHz. The Stokes ($f_S$) and anti-Stokes ($f_{AS}$) frequencies are determined from Lorentzian fits to the spectral peaks. More detail on the technique and analysis can be found in the methods section.\\

Previously, it has been shown experimentally in ultrathin films~\cite{Belmeguenai2015,Cho2015} that the effective DMI is interfacial and $D\_{eff}=D\_{s}/t\_{FM}$, where $t\_{FM}$ is the ferromagnetic thickness and $D\_{s}$ is the difference in the interfacial DMI parameter contributions from the bottom and top interfaces. The frequency shift, $\Delta f \equiv f_S -f_{AS}$, can thus be converted into the interfacial DMI constant $D_s$ using~\cite{Belmeguenai2015};
\begin{equation}
D\_{s}=D\_{eff} t\_{FM}= \frac{\pi}{2\gamma} \frac{M\_{s}t\_{FM}\Delta f}{k\_{sw}},
\end{equation}
where $\gamma$ is the gyromagnetic ratio $\gamma = g\mu\_{B}/\hbar = g \times 8.794 \times 10^{6}$~Hz/G, $t\_{FM}$ is the ferromagnetic layer thickness and $k\_{sw}$ is the SW wavenumber. As the experiments are performed in the back-scattering geometry, the scattering SW wavenumber is given by
\begin{equation}
k\_{sw} = \frac{4\pi}{\lambda\_{opt}}\sin(\theta)
\label{ksw}
\end{equation}
where $\theta$ is the angle of incidence and $\lambda\_{opt}$ is the wavelength of the illuminating light (532~nm). The linearity of $k\_{sw}$ was confirmed and the incidence angle was fixed at ${\theta= 50\ ^{\circ}}$, giving $k\_{sw}=19$~$\mu$m$^{-1}$. With a typical value for Co magnetisation, $M_s = 1400$~emu/cc and the Co g-factor $g=2.17$, $D\_{s}$ in these experimental conditions is related to $\Delta f$ by ${D\_{s} = 1.61 \times \Delta f}$, with $\Delta f$ in GHz and $D_s$ in pJ/m. \\

\begin{figure}
\centering
\includegraphics[width=\textwidth]{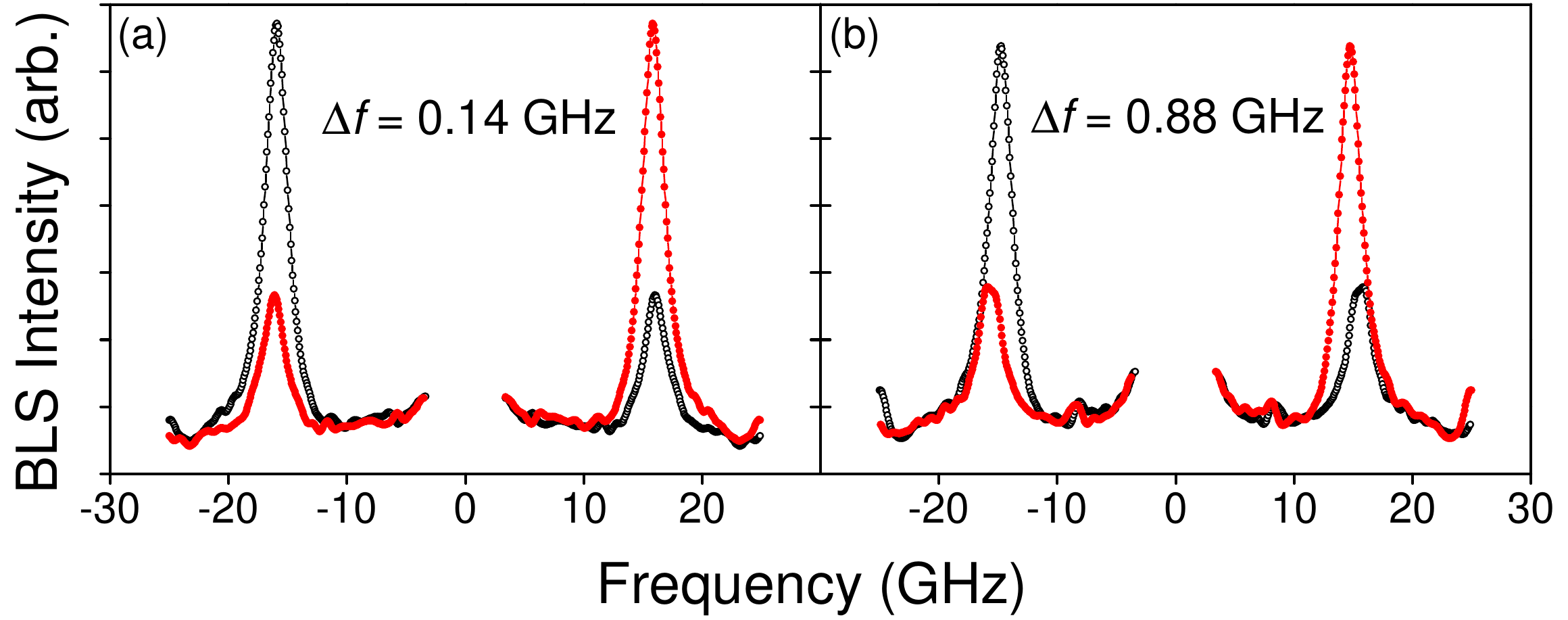}
\caption{Raw BLS spectra for (a) Pt/Co/Au (0.2 nm)/Pt and (b) Pt/Co/Au (1.6 nm)/Pt samples. Data shown by black (open) markers are obtained at +4~kOe in-plane field and at $\theta= 50 \deg$. Data in red (filled) markers show the +2~kOe field spectrum reflected in order to demonstrate the frequency shift undergone by the magnon peaks.}
\label{Fig. 1}
\end{figure}

Figure~\ref{Fig. 1} shows representative spectra for very thin (0.2~nm) and thick (1.6~nm) Au SLs. The filled markers in the figure show the spectra reflected on the horizontal axis, in order to highlight the frequency shifts more clearly. For the very thin SL, the Stokes and anti-Stokes SW lines in the direct and the frequency-inverted spectra are almost superposed ($f\_{S}\approx f\_{AS}$), indicating a very small net DMI, whereas for the thicker Au SL they show a pronounced frequency  difference, indicative of a considerable net DMI contribution. The fact that $f\_{S}<f\_{AS}$ under a positive applied field means that DMI is negative, i.e. favors left-handed cycloids \cite{Belmeguenai2015}.

\begin{figure}
\centering
\includegraphics[width=0.8\textwidth]{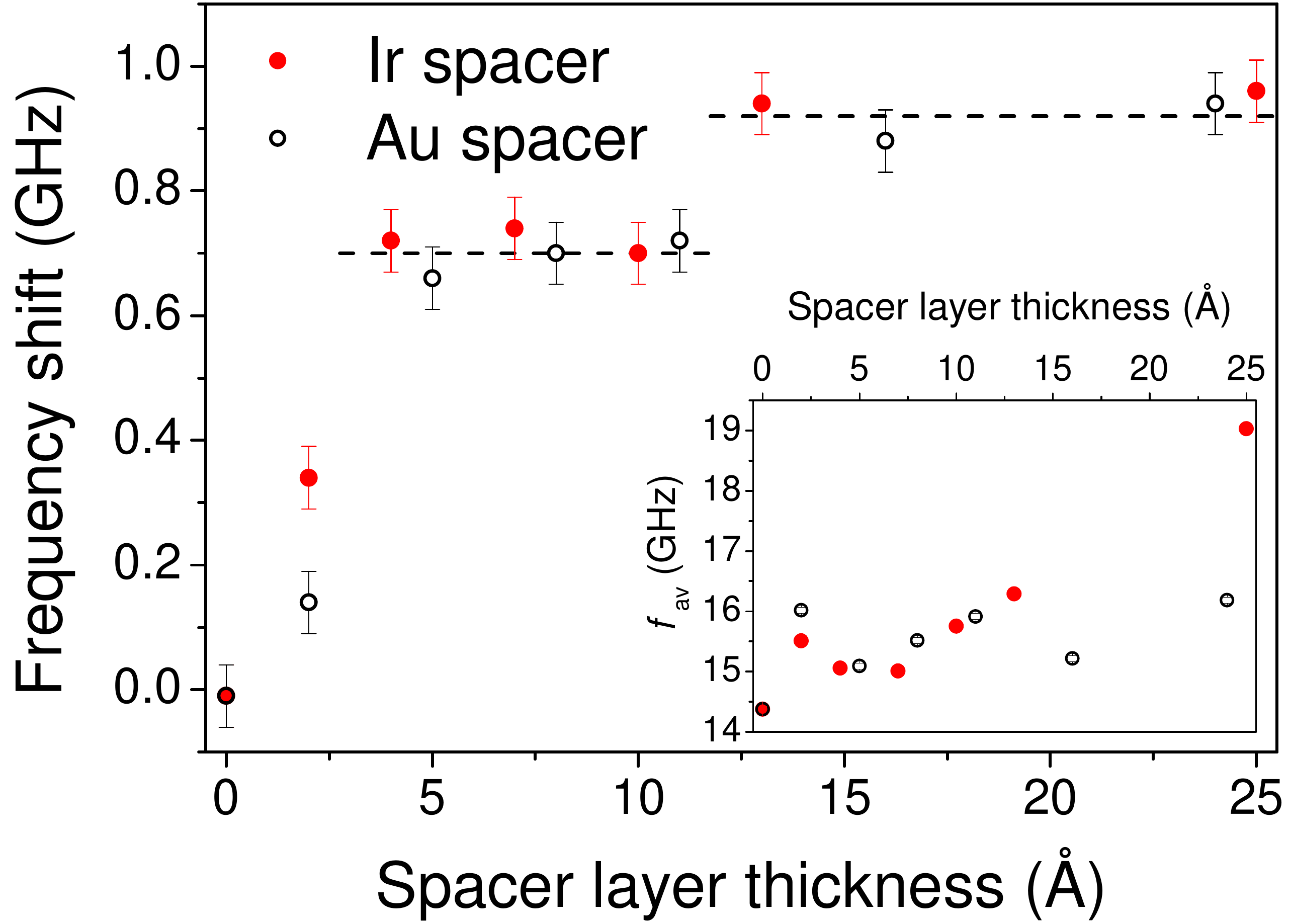}
\caption{
The frequency shift due to the DMI is plotted against nominal spacer layer thickness for Au spacers (open markers) and Ir spacers (filled markers). The rise in DMI is due to breaking the symmetry of the Pt/Co/Pt structure, removing the cancellation of the DMI contributions from the top and bottom Pt interfaces. Dashed lines indicate plateau values of the frequency shift due to DMI. Inset: Average frequency, $f_\mathrm{av} = \left(f_\mathrm{s} + f_\mathrm{as}\right)/2$, extracted from BLS, as a function of nominal spacer thickness. Error bars in $f_\mathrm{av}$ are smaller than the size of the datapoint markers.
}
\label{Fig 2}
\end{figure}

Figure ~\ref{Fig 2} summarises the change of DMI as a function of Au and Ir nominal spacer layer thickness, where the absolute value of the DMI-induced frequency asymmetry $|\Delta f|$, referred to as the frequency shift, is plotted as a function of the nominal SL thickness $t$. The first thing to note here is that with no SL, no DMI is observed. Within measurement resolution this indicates perfect cancellation of the DMI contributions from the top Co/Pt and bottom Pt/Co interfaces here. This is interesting since the Co on Pt interface is typically of better structural quality than the Pt on Co interface and elsewhere measurements of DMI, through domain expansion, by Hrabec~\etal~\cite{Hrabec2014} found that, except for epitaxial structures, the asymmetry of the microstructure between the bottom Pt/Co and the top Co/Pt interfaces gave rise to a net D$\_{s}$ in a nominally symmetric Pt/Co/Pt multilayer (it is noted that in this case the Co was much thinner and the magnetisation out of the plane, compared to the present study). A similar asymmetry was observed from BLS measurements~\cite{Belmeguenai2016} in CoFe/Pt and Pt/CoFe stacks, indicating that the structure at the Pt/Co and Co/Pt interfaces are very similar in terms of mediating the interfacial DMI.

The observed increase of the DMI with SL thickness occurs over a lengthscale that is consistent with the interfacial width, as determined by XRR analysis (see below), and is similar for both Au and Ir spacer layers. The thickness dependence shows an initially rapid rise with SL thickness and, given the inter-channel frequency separation of 82~MHz and the observed step height of 200~MHz, indicates a two-step approach to saturation for both Au and Ir spacer layers. Taking into account the finite interface width, the final step change may be associated with the formation of a continuous spacer with a complete coverage of the Co, similar to the effect shown recently for the heavy metal capping layer thickness dependence on magnetisation damping \cite{Azzawi2016}. Alternatively, the steps may reflect deviations arising from oscillations superimposed on a smooth exponential approach to saturation.\\

Within experimental precision, for both Au and Ir SLs the frequency shift saturates at $-0.96 \pm 0.01$~GHz. Assuming here that the interfacial DMI contributions of the two interfaces can be added in opposition, then the observed saturation value of -0.96~GHz corresponds to a net value, $D_\mathrm{eff}=D_{\mathrm{Pt/Co}}\mathrm{-} D_{\mathrm{Co/SL}}=-0.62$~mJ/m$^2$, which relates to a difference between the interfacial constants $D\_s$ of -1.54~pJ/m. Considering other work, no measurable DMI was observed experimentally for Au/CoNiCo/TaN \cite{Ryu2014}, while a recent Brillouin spectroscopy study~\cite{Kim2016} on Ta/Pt(Ir)/Co/AlOx measured a DMI constant for Ir/Co/AlOx that was 2.5 times smaller than for Pt/Co/AlOx, with both having the same sign. This corresponds to $D\_{s}$(Ir/Co/AlOx)$\approx -0.9$~pJ/m and $D\_{s}$(Pt/Co/AlOx) $\approx -2.2$~pJ/m. Interestingly, in a similar stack, but without a Ta seed layer, the DMI was smaller $\approx -1.4$~pJ/m, while a further study of Pt/Co/AlOx~\cite{Belmeguenai2015} gave an intermediate value of  $D\_{s}= -1.7$~pJ/m. Together these results indicate a sensitivity of the interfacial DMI to the interface structure. Also, from previous work on Pt(Ir)/Co/AlOx~\cite{Kim2016}, a value can be obtained for $D\_{s}$(Pt/Co/Ir) $\approx -1.3$~pJ/m, which is in reasonable agreement with the value obtained here, given the noted sensitivity of the DMI to the interfacial structure. \\

It was assumed here that in Ref\cite{Kim2016} the DMI contribution from the Co/AlOx interface was negligible. In contrast, for Pt/Co/Au with the assumption that $D_s$(Au/Co) $\approx 0$, as suggested in Ref~\cite{Ryu2013}, then  $D\_{s}=-2.2$~pJ/m, which is significantly larger than the $D\_{s}$(Pt/Co/Au)$\approx-1.54$~pJ/m experimental value obtained here. This discrepancy, therefore, calls into question the validity of the earlier assumption. In addition to experimental comparisons, the results here can also be compared with theoretical \textit{ab initio} calculations~\cite{Yang2015}, which gave values of $D\_{s}$ of 0.39~pJ/m for Au/Co and 0.24~pJ/m for Ir/Co interfaces respectively.
The difference between these values of $D\_{s}$ would lead to a frequency shift of only $\frac{2\gamma k_{SW}}{\pi M_s t_{FM}}(D_{Au/Co}-D_{Ir/Co}) = 72$~MHz which would be experimentally undetectable and would explain the observed similarity of the $\Delta f$ saturation values for the two SL systems.

The insert in Fig.~\ref{Fig 2}
shows the average frequencies $f\_{av} = (f\_{S} + f\_{AS})/2$ as a function of SL thickness $t$. In the absence of DMI and at $k_{sw}=0$  the resonant frequency is represented by the Kittel formula $f=(\gamma \mu_0 / 2 \pi) \sqrt{H(H+H_K-M_s)}$,
allowing any changes of the interfacial anisotropy to be tracked through changes of the effective anisotropy field, $H_K$, as a function of the spacer layer thickness. For a Au SL, only small scale variations were observed, which, at first sight, is consistent with the fact that Au/Co and Pt/Co have similar interfacial anisotropy constants \cite{denBroeder1991}, although the variation is statistically significant and suggests some relation to changes of the interface structure related to the different thermodynamics of the Co-Au and Co-Pt systems. For the Ir SL, similar behaviour was observed up to 1.8~nm SL and then an increase of frequency (decrease of anisotropy) was observed at the largest Ir thickness. The opposite trend might have been expected if only interfacial anisotropy changed, as Ir/Co has a larger interfacial anisotropy constant than Pt/Co \cite{denBroeder1991}.\\

\begin{figure}
\centering
\includegraphics[width=0.8\textwidth]{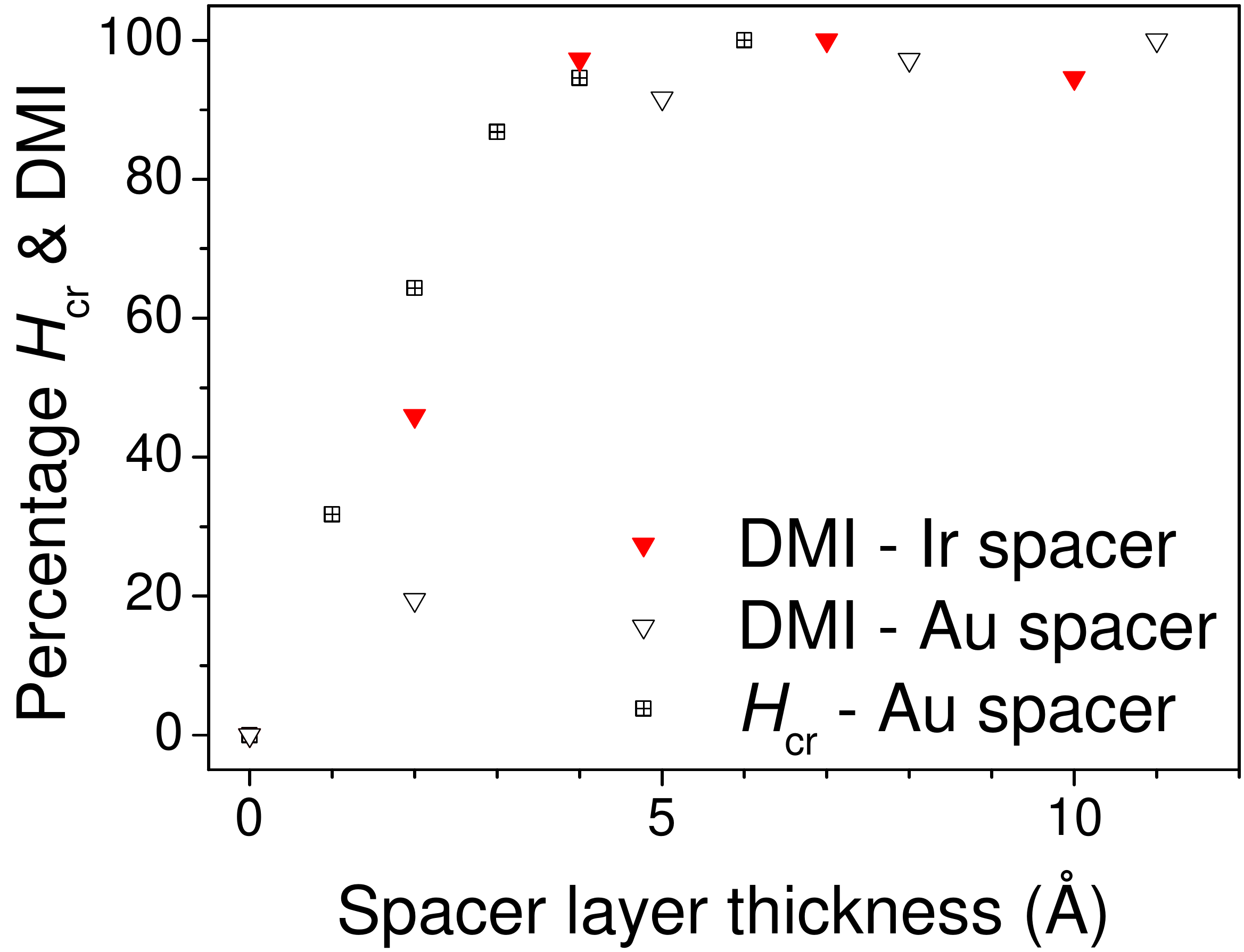}
\caption{Comparison of crossing field, $H_\mathrm{cr}$, from Ref.~\cite{Ryu2013}\ and experimental values of DMI with Au and Ir SLs from this work, across the range of comparable SL thicknesses available.  $H_\mathrm{cr}$ is normalised to the maximum value and the DMI is normalised to the first plateau value.}
\label{Fig 4}
\end{figure}

The DMI constant obtained here from the BLS studies can also be compared with the results from high-velocity domain wall (DW) dynamics~\cite{Ryu2013,Ryu2014}. The structure of the film used in those studies was Pt(5 or 1.5)/Co(0.3)/Ni(0.7)/Co(0.15)/TaN(5) and in that work a longitudinal magnetic field was applied to cancel the effective field at the DW due to the DMI and which prevented the current-induced DW motion. This has been referred to as the ``crossing field" and it is proportional to the DMI as $H_\mathrm{cr}=D_\mathrm{eff} / (\mu_0 M_s \Delta)$
where $\Delta$ is the DW width parameter. The crossing field was observed to decrease with ultrathin sub-nanometric Au spacer layers inserted at the lower Pt/Co interface. In Fig.~\ref{Fig 4} the DMI variation (expressed as a percentage) obtained here is compared to the results from Ryu~\etal. A broadly similar variation was observed for both methods for all curves for the available SL thicknesses up to 0.5~nm. In other words, estimates of the IDMI strength as a function of SL thickness from two completely different experimental techniques correlate well over the SL thicknesses available for both datasets. However, it is noted that the DMI we report here exhibits an additional step increase for thicker spacer layers, which is a non-trivial result since there exists a pronounced discrepancy between the $H_\mathrm{cr}$ measured for different heavy metal underlayers in the above cited papers~\cite{Ryu2013,Ryu2014} and the values for the interfacial DMI constants extracted from more recent independent BLS experiments~\cite{Kim2016}. Ryu~\etal~report an eight fold difference in the value of the $H_\mathrm{cr}$ in Pt/Co and Ir/Co structures whilst the BLS measurements reported in ref~\cite{Kim2016} indicate a difference that does not exceed three times. A comparative study of IDMI by similar experimental methods (DW dynamics vs BLS) showed that the measurements obtained from DW dynamics typically underestimate the IDMI strength, from 10 percent up to an order of magnitude with respect to the BLS data~\cite{Soucaille2016}.

\section*{Proximity induced magnetism as a function of Au and Ir spacer layers}
\begin{figure}
\centering
$\vcenter{\hbox{\includegraphics[width=0.48\textwidth]{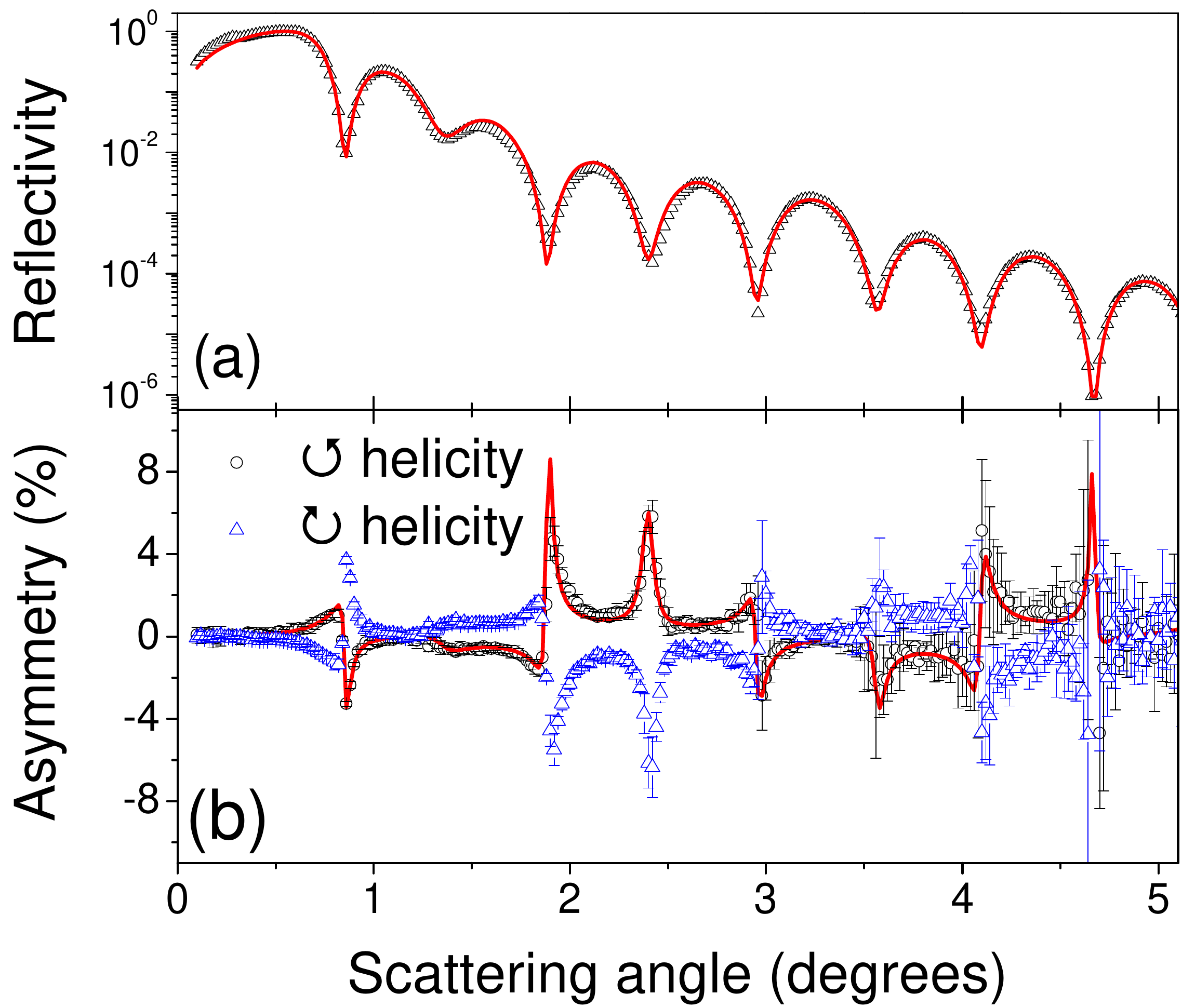}}}$
\hfill
$\vcenter{\hbox{\includegraphics[width=0.48\textwidth]{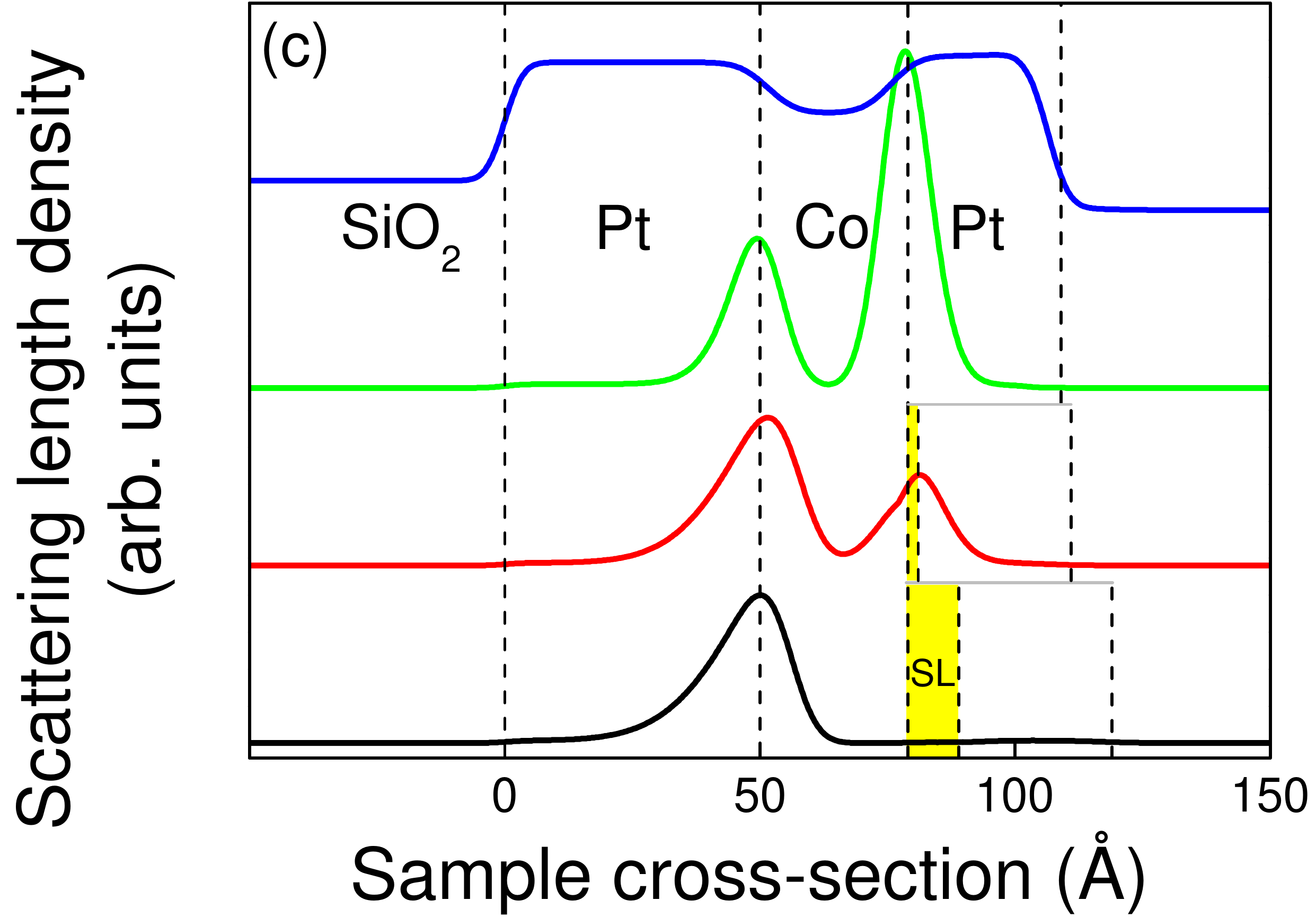}}}$
\caption{(a) Example of specular reflectivity data (markers) for no spacer layer and the best fitting simulation (line). (b) Spin asymmetry data for the same sample (markers) and best fitting simulation (line) (c) Scattering length density (SLD) profiles extracted from the best fit simulations. Upper curve structural SLD corresponding to the sample with no SL, second curve magnetic SLD with no spacer layer and lower curves corresponding mSLD with 0.2 nm and 1 nm of Au  respectively. The peaks in the mSLD coincide with the Pt interfaces (vertical dashed lines) and the area under them is related to the total moment induced at the interface. }
    \label{SLDs}
\end{figure}

Proximity induced magnetism in Pt was extracted using X-ray resonant magnetic reflectivity (XRMR)~\cite{Macke2014,Stepanov2000}, which is an element specific technique and therefore sensitive only to the Pt moment when performed using circularly polarised x-rays tuned to the energy of the Pt $L\_{3}$ absorption edge.The primary experimental quantity related to PIM from the XRMS measurements is the spin asymmetry ratio ($R\_{a}$), defined as $R\_{a} = (I^{+}-I^{-})/(I^{+}+I^{-})$ where $I^{+}$ is the scattered intensity of the circularly polarized x-rays when the sample is magnetized with a positive in-plane magnetic field and $I^{-}$ is the equivalent with negative magnetic field. The reflectivity geometry allows the depth dependence of the PIM to be extracted from  simultaneous best fitting simulations of the measured specular reflectivity, $R\_{s} = (I^{+}+I^{-})/2$, (Fig.~\ref{SLDs}(a)) and the spin asymmetry ratio  (Fig.~\ref{SLDs}(b)), thereby constraining the best fitting magnetisation profile to be consistent with the physical structure of the sample. For the simulations the sample was modeled as slabs of Pt, Co and a SL of various thickness and interfacial roughness. From the model, the structural scattering length density profile (sSLD) was extracted, an example of which is shown in the upper trace of Fig.~\ref{SLDs}(c). The spin asymmetry ratio was fitted by modeling the sample structure with a distribution of magnetic moments through the thickness of the Pt, which gave the magnetic scattering length density profile (mSLD), examples of which are shown in the lower traces of Fig.~\ref{SLDs}(c).  More detail on the technique and analysis can be found in the methods section.\\

Fig.~\ref{SLDs}(c) shows the sSLD that represents the structure of the Pt/Co/Pt sample. The plateaux correspond to the bottom Pt and top Pt layers separated by the lower (non-resonant) sSLD of the Co and the slopes between indicate the interface widths. The corresponding mSLD for Pt/Co/Pt is shown immediately below the sSLD. A peak occurs at each Pt/Co interface, indicated by the vertical dashed lines. The area under each peak is proportional to the total induced Pt moment at that Pt interface. The lower Pt/Co interface is the same for all samples, so the induced moment can be assumed constant, such that for samples with increasing SL thickness, any variations of experimental conditions between samples is accounted for by normalizing the area under the Co/SL/Pt peak to the area under the lower Pt/Co layer mSLD peak, as a function of SL. Considering now the details of the induced Pt moments at the upper and lower interfaces. We first, note that the induced Pt moment for the sample with no SL shows a larger moment for the top Co/Pt interface compared to the lower Pt/Co interface. Such asymmetries of the induced moment have also recently been observed elsewhere in Pd/Co/Pd structures~\cite{Kim2016a}, suggesting differences in the interfacial structures at the top and bottom interfaces of the Co layer that result from the growth of Co onto Pt and Pt onto Co respectively. A tomographic atomic probe study by Lard\'{e}~\etal~\cite{Larde2009} using Pt/Co multilayers did find a difference in the composition profile across the Pt/Co interface relative to the Co/Pt interface. However, no evidence of different interfacial structures was observed in the best fits to the specular reflectivity, which may reflect the subtlety of any differences in interfacial structure. Nonetheless, the result suggests that the proximity induced magnetisation of Pt is very sensitive to the local atomic structure and associated electronic interactions.\\

Interestingly, the difference in induced moment between the top and bottom Pt interfaces is not reflected in the DMI measurement, which within the measurement resolution showed perfect cancellation of the DMI contributions. Since the DMI is also sensitive to the structure at the interface, this suggests that the different sensitivities to the interface structure of PIM and the DMI argues against the requirement for a direct linkage between PIM and DMI~\cite{Ryu2014}. However, if the difference in the Pt moments induced at the bottom and top interfaces was due to different degrees of interfacial intermixing, not resolved by the specular reflectivity measurements, this would also contradict the~\textit{ab initio} calculations of Yang~\etal~\cite{Yang2015} that suggested intermixing has a large detrimental effect on the DMI. Thus, it me be concluded that local differences in interfacial structure may contribute to the differences in the proximity induced moments.\\

With the addition of a sub-nanometric SL, the mSLD changes significantly, with the moment on the top Pt layer falling dramatically for both Au and Ir spacer layers. As shown in Fig.~\ref{SLDs}(c), The insertion of a nominal 0.2~nm layer of Au significantly changes the magnetic profile of the top Pt layer, reducing the induced Pt moment at the top interface to below that of the bottom interface. This further demonstrates the exquisite sensitivity of the PIM to the structure of the interface, in particular to the immediate proximity of Pt to Co. The addition of 1~nm of Au between the layers almost entirely destroys the induced magnetism in the upper Pt layer.\\

\begin{figure}
\centering
\includegraphics[width=0.75\textwidth]{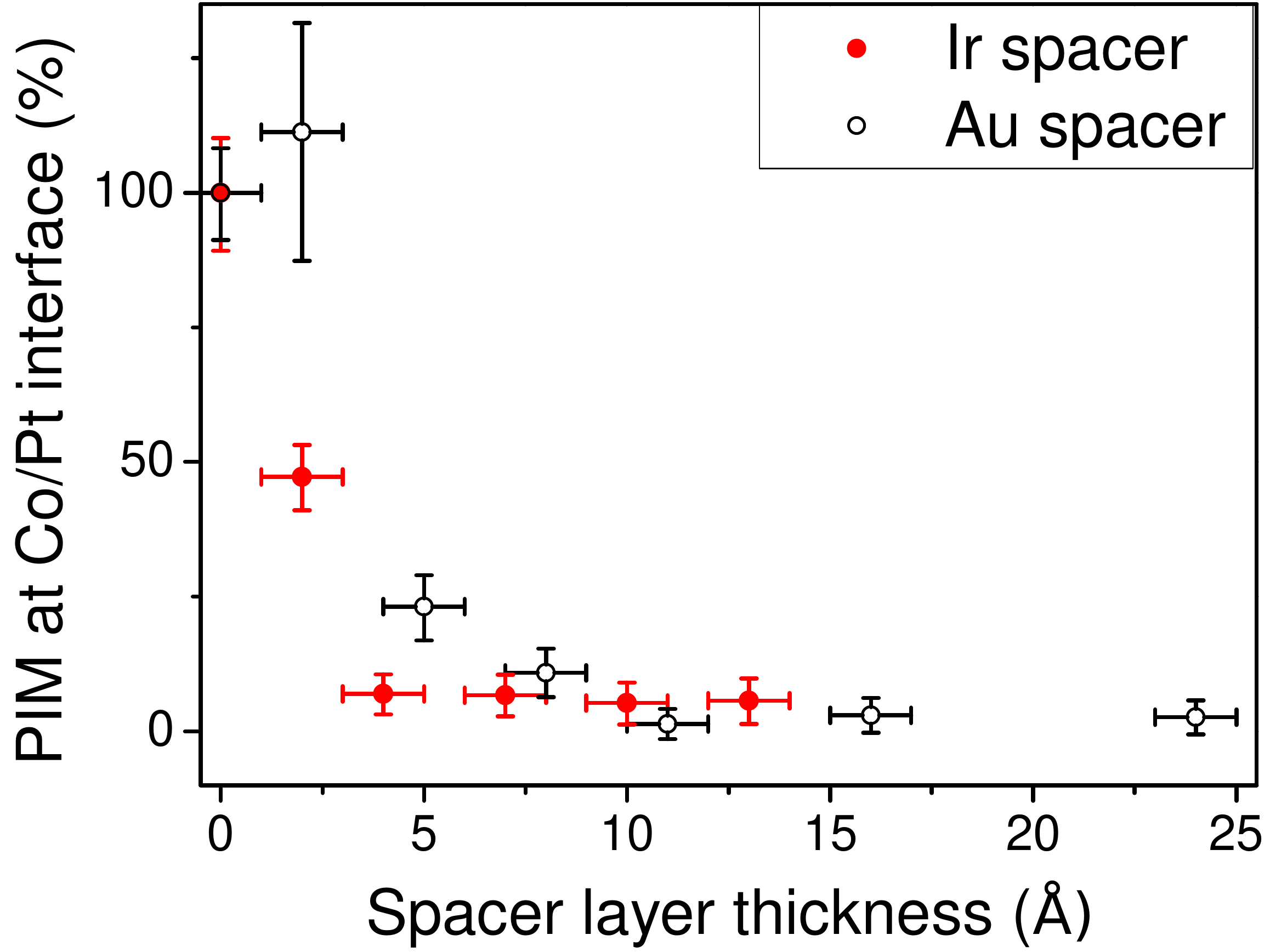}
\caption{Percentage of Pt moment at top interface  as a function of Au or Ir nominal spacer layer thickness. The Pt moment falls rapidly with increasing spacer layer thickness.}
\label{Ptperc}
\end{figure}

The PIM as a function of Au and Ir spacers is summarised in  Fig.~\ref{Ptperc}, where the ratio of the areas under mSLD peaks of the upper Pt and lower (fixed) Pt is plotted as a function of the SL thickness, and normalised such that the sample with no spacer layer has a value of 100~\%. This shows the percentage moment remaining in the top Pt layer relative to the zero spacer layer sample. For the Ir spacer the PIM in the Pt layer is completely lost (within experimental error) for a SL thickness of 0.4-0.7~nm, whereas with Au SL the Pt PIM is lost with a SL thickness of 1.1~nm. Thus Ir seems to cause the PIM at the top Co/Pt interface to reduce more rapidly, while the Au thickness dependence is a little more gradual.
The indication that Ir causes a slightly more rapid loss of the Pt PIM may be considered surprising since Ir can take on a proximity induced moment~\cite{Wilhelm2001,Stahler1993}, which may have been considered to mediate any polarising interaction between the Co and the Pt.\\

When considering the role of ultrathin SLs the nature of the interface is important. From analysis of the specular x-ray reflectivity the interface width of the Co/Pt interface was determined to be of order 0.5~nm, which can be considered simplistically as a guide to the minimum SL thickness required to create a continuous layer at the interface.The difference between Au and Ir is their ability to alloy with Pt. Au is almost entirely immiscible with Pt~\cite{Okamoto1985}, so sub-nanometric SLs of Au would form discontinuous layers allowing a significant amount of direct contact between Co and Pt at the interface. In contrast, Ir is expected to alloy with Pt~\cite{Bharadwaj1995}, so an Ir SL would locally intermix with Pt, alloying Ir and Pt at the interface, which would cause significant modification to the density of states of the Pt. The interfacial Pt may then become less strongly proximity polarised as the Stoner enhanced paramagnetic susceptibility would be reduced from that of pure Pt. For ultrathin Au SL the coverage would lead to a greater roughness with minimal intermixing at the SL/Pt interface, compared to a more intermixed interface with lower topological roughness for the Ir SL. These structural interface components cannot be resolved using specular reflectivity. So, the PIM dependence on the nominal SL thickness can also be related to the interfacial structure; where the local interfacial Pt-Ir alloying leads to a more rapid loss of PIM with nominal SL thickness than does the more gradual reduction in direct Co/Pt contact caused by the insertion of a sub-monolayer Au SL.

\section*{Relationship between DMI and Proximity induced magnetisation in Pt }
The characteristic features can be summed up briefly as follows, both DMI and PIM  demonstrate a rapid approach to saturation as a function of SL thickness, but they do not follow the same trend. The PIM falls rapidly with both Au and Ir SLs. For Ir, a nominal thickness greater between 0.4 and 0.7~nm is required for the complete loss of the Pt moment, which coincides with the thickness required for complete sign reversal of the DMI-generated in-plane magnetic field, $H\_{DMI}$, observed for perpendicular Pt/Co/Ir/Pt multilayers in Ref.~\cite{Hrabec2014}.  The DMI rises rapidly with the initial insertion of a SL, but the approach to saturation with thicker SLs occurs via two step process, which is not reflected in the trend of the PIM data.\\

\begin{figure}
\centering
\includegraphics[width=0.75\textwidth]{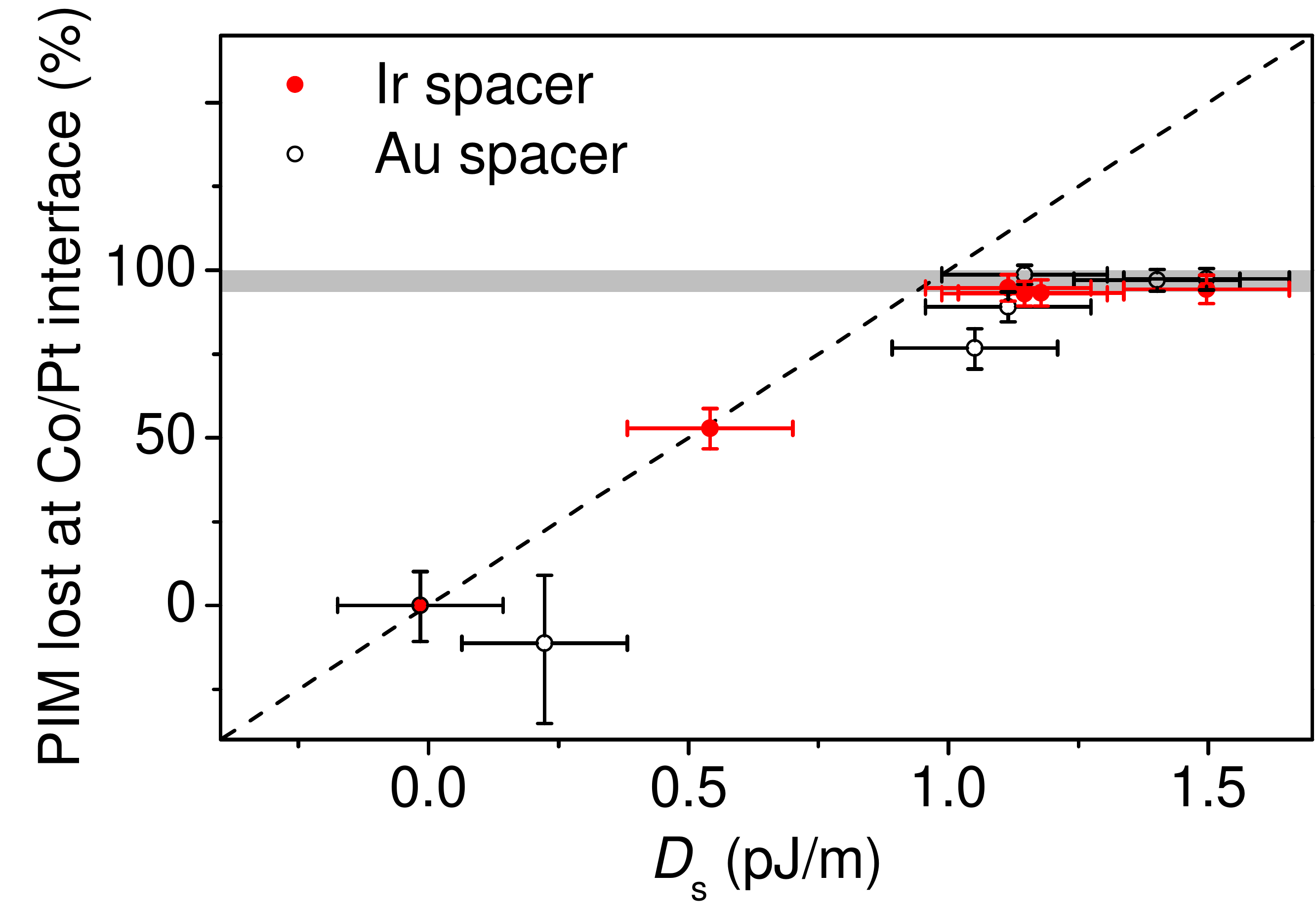}
\caption{The percentage proximity induced moment lost at the top Co/Pt interface plotted against the $D\_{s}$ for  Au and Ir SLs. The dashed line denotes a 1:1 correlation. The grey bar denotes the region where the PIM is considered to have vanished, based on the sensitivity of the XRMR measurement.
}
\label{PIM_DMI}
\end{figure}

To further investigate potential correlations between the DMI and PIM, in Fig.~\ref{PIM_DMI} the percentage of the Pt proximity moment lost from the top interface is plotted against the effective net DMI constant, $D\_{s}$. The dashed line in fig.~\ref{PIM_DMI} represents a 1:1 correspondence. For the thinnest nominal SL thickness the trend is linear, although the number of datapoints in the low DMI region is somewhat limited. Subsequent data fall below the dashed line, indicating that the PIM is lost more quickly than the DMI changes, as a function of increasing SL thickness and the data also shows that the DMI continues to rise as the SL thickness increases for thicknesses where the induced moment is completely absent.\\

The PIM is directly related to the Stoner factor, while the DMI is directly linked to the spin-orbit interaction (SOI). Theory has so far proposed a number of different realisations of the anisotropic exchange interaction. While Dzyaloshinskii~\cite{Dzyaloshinskii1958} was the first to predict it from purely symmetry grounds, it was  Moriya~\cite{Moriya1960} who suggested that low symmetry dielectrics it can be seen as a combined effect of SOI and exchange interaction. His formalism is essentially an extension of the superexchange theory to include the effect of SOI, in other words the key role is played by the “$d$-$d$” hybridization of core electrons (“$d$-$d$” exchange mechanism). A completely different mechanism was put forward to explain peculiar magnetic behaviour of spin glasses with nonmagnetic transition-metal impurities by Fert and Levy~\cite{Fert1980}. The experimentally observed enhancement of the anisotropy field can arise from an additional term in the Ruderman-Kittel-Kasuya-Yosida (RKKY) interaction of the Dzyaloshinskii-Moriya type and is due to spin-orbit scattering of the conduction electrons by the nonmagnetic impurities, which implies an active role must be played by itinerant “$s$” electrons (“$s$-$d$” exchange mechanism). In other words, the conduction electrons experience the strong spin-orbit forces of the $d$ states, not unlike the SHE. It is this mechanism that seems more relevant to the metallic structures investigated in this study. Note that whatever the specific exchange coupling mechanism, the role of the SOI is essential, in other words there can be no DMI without SOI.
At the same time, the direct short-ranged interfacial $d$-$d$ orbital hybridisation between the Co and Pt should not be overlooked either. It is instructive to compare the two mechanisms; hybridization of 
$d$-electrons and RKKY interactions via itinerant 
electrons.
The unmistakable signature of the conduction electron dominated RKKY interaction is oscillatory behaviour as a function of spacer thickness $t$ with a characteristic period $\Lambda = \frac{\lambda_{F}}2$, where $\lambda_{F}$ is the Fermi wavelength~\cite{Balten1990}. To the first approximation, it asymptotically decreases as the reciprocal square of thickness. In Au $\lambda_{F}$  is on  the order of 0.5~nm, thus $\Lambda \approx 0.25$~nm, which makessuch short-period oscillations practically undetectable experimentally . What can actually be observed in an experiment are multiperiodic long-range ($\approx 1$~nm) oscillations~\cite{Purcell1992,Johnson1992,Fuss1992} that occur due to the discrete character of the atomic structure and a beat phenomenon arising from the incommensurability of $t$ and inter-atomic distance~\cite{Bruno1992}. In this regard, the step-like irregularities featured by the DMI dependence on SL thickness curve bear a resemblance with the ripple produced by RKKY conduction electron interferences and thus attest to relevance of the itinerant electrons to the DMI related mechanisms. The oscillatory nature of the RKKY-based DMI mechanism in metals appears directly in the theoretical description proposed by Fert~\cite{Fert1980}. No such features occur in the PIM SL thickness dependences, which is not surprising since the RKKY mechanism is not associated with the PIM. In other words, the differences in the evolution of the DMI and PIM with increasing spacer thickness suggests a predominant role played by the itinerant electrons in the case of DMI and bounded $d$-electrons in the case of the PIM.\\

As for the absolute values, no explanation of the link between the two phenomena can be proposed, unless there exists some subtle implicit correlation between the symmetries simultaneously favouring the SOI and high Stoner factor. Interestingly, recent numerical simulations~\cite{Yang2015} lead to an opposite conclusion, namely the more the spin configuration is favourable for the DMI, the less the PIM is pronounced and vice versa. It should also be mentioned that the conventional interpretation of PIM as an extension of ferromagnetic state from a FM into an adjacent heavy metal, typically  paramagnetic, through their common interface, is somewhat simplistic.

\section*{Conclusions}

PIM and DMI at the Co/Pt  interface have been investigated in detail as a function of spacer layers of Au and Ir of various thickness. The SL thickness dependence of both DMI and PIM demonstrate a rapid change as a function of SL thickness, but not in the same way. First, there exists a major difference in the nominally symmetric Pt/Co/Pt trilayer without a SL. In this structure the proximity induced magnetism manifests a pronounced asymmetry, the PIM developed within the top Pt layer is at least twice  that of the bottom one. In contrast, the interfacial Dzyaloshinskii-Moriya Interaction in the same Pt/Co/Pt system is characterized  by a perfectly symmetric spatial pattern, indicating identical DMI strength at both Pt/Co and Co/Pt interfaces, which leads to a complete cancellation of both contributions, i.e. a zero net DMI effect.
Secondly, for both Au and Ir SLs the PIM decays rapidly in a regular monotonic fashion, vanishing for SL thickness greater than about 1.0~nm. For Ir the decay is slightly more rapid, with all PIM destroyed by sL thickness of 0.4-0.7~nm. The results indicate the loss of proximity induced magnetisation in Pt occurs on a shorter lengthscale for Ir than for a Au spacer layer.
Similarly, the observed increase of the DMI with SL thickness occurs on a lengthscale that is consistent with the interfacial width, as determined by XRR. But, the lengthscale is similar for either Au or Ir spacer layers. For both Au and Ir spacer layers the net DMI saturates to the same value of $D\_{s} = -1.54$~pJ/m. Moreover, the approach to saturation takes place in a characteristic two-step manner.
This variation may be associated with the formation of a spacer with a complete coverage of the Co. Alternatively, this feature bears a resemblance with the oscillation produced by RKKY conduction electron interferences cautiously suggesting the relevance of the itinerant $s$-electrons to the DMI related mechanisms.


\section*{Methods}
\subsection*{Analysis of X-ray resonant magnetic reflectivity}
When simulating the sample, to accommodate the thin spacer layers the sample is sliced into 0.05~nm sections. A scattering length density is calculated for each of these sections. This is calculated as the product of the fitted electron density for that particular slice and the x-ray scattering factor for the material, in units of Thompson scattering lengths per cubic angstrom.

The x-ray scattering factors can be written as~\cite{Bjorck2014},
\begin{equation}
f(Q,E) = (\hat{\epsilon}_{f}\cdot\hat{\epsilon}_{i})F^{0}(E) - i(\hat{\epsilon}_{f}\times\hat{\epsilon}_{i})\cdot \hat{m}F^{1}(E).
\end{equation}
Here, $Q$ and $E$ are the scattering vector and x-ray energy respectively. $\hat{\epsilon}\_{f}$ and $\hat{\epsilon}$ are unit vectors describing the polarisation state of the incident and scattered x-rays. $F^{0}$ is the charge scattering amplitude, comprised of both real $f\_{0} + f'(E)$ and imaginary $f''(E)$. Where $f_{0}$ is the Thompson scattering factor, roughly equal to the atomic number of the element with $f'$ and $f''$ the anomalous dispersive corrections due to the energy dependence cloe to resonance. Finally, $F^{1}$ is the equivalent magnetic scattering amplitude, containing analogous real $m'$ and imaginary $m''$ components which account for the magnetisation and polarisation dependence.

The separate slices are then arranged alongside each other to produce a profile through the sample. This profile is then used to calculate the Fresnel coefficients necessary for the Parratt recursive method employed by the model. The parameters are then fitted using a differential evolution algorithm~\cite{Bjorck2011} which adjusts the SLD profiles until a good fit is found for the data.

Due to the large number of parameters used to fit a given model, it is essential to  constrain the fit. We have achieved this by simultaneously fitting the reflectivity alongside the spin asymmetry, thereby ensuring all the structural parameters are consistent with both datasets. Furthermore the resonant scattering corrections, $f'$ and $f''$ as well as the magnetic scattering factors $m'$ and $m''$ , which describe the resonant behaviour of the Pt, were determined from the sample with no SL and fixed for the other samples in the series. This approach allows for direct comparisons to made between the samples, but means an absolute magnetic moment can not be established. However, it is the relative change in magnetism as a function of SL that we are interested in.

\subsection*{Experimental procedure and analysis of BLS spectra}

To ensure reliable Stokes/anti-Stokes frequency asymmetry detection, both the experimental procedure and processing of BLS spectra have been arranged accordingly.
On the experimental level the main risk lies with the zero frequency shift in the BLS spectra. This is especially important in the case of ultrathin SLs, where the net DMI tends to zero through the mutual cancellation of the two contributions from opposite interfaces and, consequently, the Stokes/anti-Stokes frequency asymmetry is very small. Thus, to minimize the instrumental error the following complementary measurements have been taken. First, we were periodically taking BLS spectra without an analyser, thus recording not only the peaks corresponding to light scattering by magnons, but also those representing phonons. Importantly, the propagation of acoustic phonons is never non-reciprocal, hence $\Delta f \equiv f_S - f_{AS}=0$  which makes them a very effective instrument for checking the calibration of the frequency sweeping of the BLS set-up.
This is illustrated in Supplementary Fig.\ref{figmethod} which shows a representative spectrum (red solid line), obtained at an incidence angle $\theta = 50 ^\circ$ ($k_{SW} = 19$~$\mu\mathrm{m}^{-1}$) and without analyser such that the phonon lines could also be represented in the spectrum. The spectrum in blue dotted line is the BLS distribution reflected in the vertical axis in order to show the frequency shifts more clearly. It contains three peaks for negative (Stokes) and positive (anti-Stokes) frequency shifts. The two low-frequency peaks around 10~GHz that do not shift with field and disappear in the crossed polarisers configuration are produced by phonons while the third high-frequency one at ~$\pm 23$~GHz, whose frequency position is directly controlled with an external magnetic field is due to BLS from magnons. As can be seen in Supplementary Fig. \ref{figmethod} both phonon spectra are absolutely symmetric, the Stokes and anti-Stokes lines in the direct (solid red) and frequency-inverted (dotted blue) spectra are perfectly overlapping ($f_S=f_{AS}$), whilst the magnon-related line demonstrates a clearly seen asymmetry ($f_S < f_{AS}$). It was routine in measurements to place the analyser in front of the Fabry-Perot interferometer to improve the signal to noise ratio. Moreover, to double check it, we were taking advantage of the fact that the sign of delta is changed if the saturating magnetic field polarity is inversed. Spectra were obtained after counting photons for typically 12 hours to improve the quality of the raw data BLS spectra.

Numerical simulations of the shape of the BLS spectra of SW in the DE geometry were very helpful for improving the accuracy and reliability of our measurements. They were performed with the help of an \textit{ad hoc} code based on the fluctuation dissipation theorem to evaluate the thermally activated SW and the Green functions associated with the light scattering~\cite{CamleyMills}. For the analysis contained herein we have introduced appropriate boundary conditions \cite{Kost2014} to calculate spectra in the case of a ferromagnetic film between two layers having different surface anisotropies and interface DMI constants. These calculations provide the positions as well as the heights of the SW lines. In particular, they have have permitted estimation of the contribution to $\Delta f$ of the conventional one-sided surface magnetic anisotropy which has turned out to be under 0.01~GHz, in other words far too small to undermine the reliability of our observations. On the other hand, the above-mentioned code has proven to be effective for optimizing the cross-section of the BLS by thermal magnon, which is very sensitive to phase relations between numerous optical reflections in a multilayer system ensuring their constructive interferences.


\bibliography{PIM_BLSpaper}

\begin{thebibliography}{10}
\expandafter\ifx\csname url\endcsname\relax
  \def\url#1{\texttt{#1}}\fi
\expandafter\ifx\csname urlprefix\endcsname\relax\def\urlprefix{URL }\fi
\providecommand{\bibinfo}[2]{#2}
\providecommand{\eprint}[2][]{\url{#2}}

\bibitem{Sankey2008}
\bibinfo{author}{Sankey, J.~C.} \emph{et~al.}
\newblock \bibinfo{title}{Measurement of the spin-transfer-torque vector in
  magnetic tunnel junctions}.
\newblock \emph{\bibinfo{journal}{Nat Phys}} \textbf{\bibinfo{volume}{4}},
  \bibinfo{pages}{67--71} (\bibinfo{year}{2008}).

\bibitem{Parkin2008}
\bibinfo{author}{Parkin, S. S.~P.}, \bibinfo{author}{Hayashi, M.} \&
  \bibinfo{author}{Thomas, L.}
\newblock \bibinfo{title}{Magnetic domain-wall racetrack memory}.
\newblock \emph{\bibinfo{journal}{Science}} \textbf{\bibinfo{volume}{320}},
  \bibinfo{pages}{190--194} (\bibinfo{year}{2008}).

\bibitem{Rashba1984}
\bibinfo{author}{Bychkov, Y.~A.} \& \bibinfo{author}{Rashba, E.~I.}
\newblock \bibinfo{title}{Properties of a 2d electron gas with lifted spectral
  degeneracy}.
\newblock \emph{\bibinfo{journal}{J. Exp. Theor. Phys. Lett.}}
  \textbf{\bibinfo{volume}{39}}, \bibinfo{pages}{78--81}
  (\bibinfo{year}{1984}).

\bibitem{MihaiMiron2010}
\bibinfo{author}{Mihai~Miron, I.} \emph{et~al.}
\newblock \bibinfo{title}{Current-driven spin torque induced by the rashba
  effect in a ferromagnetic metal layer}.
\newblock \emph{\bibinfo{journal}{Nat Mater}} \textbf{\bibinfo{volume}{9}},
  \bibinfo{pages}{230--234} (\bibinfo{year}{2010}).

\bibitem{Hirsch1999}
\bibinfo{author}{Hirsch, J.~E.}
\newblock \bibinfo{title}{Spin hall effect}.
\newblock \emph{\bibinfo{journal}{Phys. Rev. Lett.}}
  \textbf{\bibinfo{volume}{83}}, \bibinfo{pages}{1834--1837}
  (\bibinfo{year}{1999}).

\bibitem{Ganguly2014}
\bibinfo{author}{Ganguly, A.} \emph{et~al.}
\newblock \bibinfo{title}{Time-domain detection of current controlled
  magnetization damping in pt/ni81fe19 bilayer and determination of pt spin
  hall angle}.
\newblock \emph{\bibinfo{journal}{Applied Physics Letters}}
  \textbf{\bibinfo{volume}{105}}, \bibinfo{pages}{--} (\bibinfo{year}{2014}).

\bibitem{Emori2012}
\bibinfo{author}{Emori, S.}, \bibinfo{author}{Bono, D.~C.} \&
  \bibinfo{author}{Beach, G. S.~D.}
\newblock \bibinfo{title}{Interfacial current-induced torques in pt/co/gdox}.
\newblock \emph{\bibinfo{journal}{Applied Physics Letters}}
  \textbf{\bibinfo{volume}{101}}, \bibinfo{pages}{--} (\bibinfo{year}{2012}).

\bibitem{Schellekens2012}
\bibinfo{author}{Schellekens, A.}, \bibinfo{author}{van~den Brink, A.},
  \bibinfo{author}{Franken, J.}, \bibinfo{author}{Swagten, H.} \&
  \bibinfo{author}{Koopmans, B.}
\newblock \bibinfo{title}{Electric-field control of domain wall motion in
  perpendicularly magnetized materials}.
\newblock \emph{\bibinfo{journal}{Nat Commun}} \textbf{\bibinfo{volume}{3}},
  \bibinfo{pages}{847--} (\bibinfo{year}{2012}).

\bibitem{Ryu2012}
\bibinfo{author}{Ryu, K.-S.}, \bibinfo{author}{Thomas, L.},
  \bibinfo{author}{Yang, S.-H.} \& \bibinfo{author}{Parkin, S. S.~P.}
\newblock \bibinfo{title}{Current induced tilting of domain walls in high
  velocity motion along perpendicularly magnetized micron-sized co/ni/co
  racetracks}.
\newblock \emph{\bibinfo{journal}{Applied Physics Express}}
  \textbf{\bibinfo{volume}{5}}, \bibinfo{pages}{093006} (\bibinfo{year}{2012}).

\bibitem{Haazen2013}
\bibinfo{author}{Haazen, P. P.~J.} \emph{et~al.}
\newblock \bibinfo{title}{Domain wall depinning governed by the spin hall
  effect}.
\newblock \emph{\bibinfo{journal}{Nat Mater}} \textbf{\bibinfo{volume}{12}},
  \bibinfo{pages}{299--303} (\bibinfo{year}{2013}).

\bibitem{Emori2013}
\bibinfo{author}{Emori, S.}, \bibinfo{author}{Bauer, U.}, \bibinfo{author}{Ahn,
  S.-M.}, \bibinfo{author}{Martinez, E.} \& \bibinfo{author}{Beach, G. S.~D.}
\newblock \bibinfo{title}{Current-driven dynamics of chiral ferromagnetic
  domain walls}.
\newblock \emph{\bibinfo{journal}{Nat Mater}} \textbf{\bibinfo{volume}{12}},
  \bibinfo{pages}{611--616} (\bibinfo{year}{2013}).

\bibitem{Dzyaloshinskii1958}
\bibinfo{author}{Dzyaloshinskii, I.~E.}
\newblock \bibinfo{title}{A thermodynamic theory of weak ferromagnetism of
  antiferromagnetics}.
\newblock \emph{\bibinfo{journal}{Journal of Physics and Chemistry of Solids}}
  \textbf{\bibinfo{volume}{4}}, \bibinfo{pages}{241--255}
  (\bibinfo{year}{1958}).

\bibitem{Moriya1960}
\bibinfo{author}{Moriya, T.}
\newblock \bibinfo{title}{New mechanism of anisotropic superexchange
  interaction}.
\newblock \emph{\bibinfo{journal}{Phys. Rev. Lett.}}
  \textbf{\bibinfo{volume}{4}}, \bibinfo{pages}{228--230}
  (\bibinfo{year}{1960}).

\bibitem{Thiaville2012}
\bibinfo{author}{Thiaville, A.}, \bibinfo{author}{Rohart, S.},
  \bibinfo{author}{Ju\'{e}, E.}, \bibinfo{author}{Cros, V.} \&
  \bibinfo{author}{Fert, A.}
\newblock \bibinfo{title}{Dynamics of dzyaloshinskii domain walls in ultrathin
  magnetic films}.
\newblock \emph{\bibinfo{journal}{EPL (Europhysics Letters)}}
  \textbf{\bibinfo{volume}{100}}, \bibinfo{pages}{57002}
  (\bibinfo{year}{2012}).

\bibitem{Yu2010}
\bibinfo{author}{Yu, X.~Z.} \emph{et~al.}
\newblock \bibinfo{title}{Real-space observation of a two-dimensional skyrmion
  crystal}.
\newblock \emph{\bibinfo{journal}{Nature}} \textbf{\bibinfo{volume}{465}},
  \bibinfo{pages}{901--904} (\bibinfo{year}{2010}).

\bibitem{Heinze2011}
\bibinfo{author}{Heinze, S.} \emph{et~al.}
\newblock \bibinfo{title}{Spontaneous atomic-scale magnetic skyrmion lattice in
  two dimensions}.
\newblock \emph{\bibinfo{journal}{Nat Phys}} \textbf{\bibinfo{volume}{7}},
  \bibinfo{pages}{713--718} (\bibinfo{year}{2011}).

\bibitem{Boulle2016}
\bibinfo{author}{Boulle, O.}, \bibinfo{author}{Vogel, J.} \&
  \bibinfo{author}{Stashkevich, A.}
\newblock \emph{\bibinfo{journal}{Nature Nanotech}}  (\bibinfo{year}{2016}).

\bibitem{Geissler2001}
\bibinfo{author}{Geissler, J.} \emph{et~al.}
\newblock \bibinfo{title}{Pt magnetization profile in a pt/co bilayer studied
  by resonant magnetic x-ray reflectometry}.
\newblock \emph{\bibinfo{journal}{Phys. Rev. B}} \textbf{\bibinfo{volume}{65}},
  \bibinfo{pages}{020405--} (\bibinfo{year}{2001}).

\bibitem{Clogston1962}
\bibinfo{author}{Clogston, A.~M.} \emph{et~al.}
\newblock \bibinfo{title}{Local magnetic moment associated with an iron atom
  dissolved in various transition metal alloys}.
\newblock \emph{\bibinfo{journal}{Phys. Rev.}} \textbf{\bibinfo{volume}{125}},
  \bibinfo{pages}{541--552} (\bibinfo{year}{1962}).

\bibitem{Hase2014}
\bibinfo{author}{Hase, T. P.~A.} \emph{et~al.}
\newblock \bibinfo{title}{Proximity effects on dimensionality and magnetic
  ordering in pd/fe/pd trialyers}.
\newblock \emph{\bibinfo{journal}{Phys. Rev. B}} \textbf{\bibinfo{volume}{90}},
  \bibinfo{pages}{104403--} (\bibinfo{year}{2014}).

\bibitem{Wilhelm2001}
\bibinfo{author}{Wilhelm, F.} \emph{et~al.}
\newblock \bibinfo{title}{Systematics of the induced magnetic moments in
  $5\mathit{d}$ layers and the violation of the third hund's rule}.
\newblock \emph{\bibinfo{journal}{Phys. Rev. Lett.}}
  \textbf{\bibinfo{volume}{87}}, \bibinfo{pages}{207202--}
  (\bibinfo{year}{2001}).

\bibitem{Stahler1993}
\bibinfo{author}{Schutz, G.} \emph{et~al.}
\newblock \bibinfo{title}{Distribution of magnetic moments in co/pt and
  co/pt/ir/pt multilayers detected by magnetic x-ray absorption}.
\newblock \emph{\bibinfo{journal}{J. Appl. Phys.}}
  \textbf{\bibinfo{volume}{73}}, \bibinfo{pages}{6430--6432}
  (\bibinfo{year}{1993}).

\bibitem{Jamali2013}
\bibinfo{author}{Jamali, M.} \emph{et~al.}
\newblock \bibinfo{title}{Spin-orbit torques in co/pd multilayer nanowires}.
\newblock \emph{\bibinfo{journal}{Phys. Rev. Lett.}}
  \textbf{\bibinfo{volume}{111}}, \bibinfo{pages}{246602--}
  (\bibinfo{year}{2013}).

\bibitem{Martinez2013}
\bibinfo{author}{Martinez, E.}, \bibinfo{author}{Emori, S.} \&
  \bibinfo{author}{Beach, G. S.~D.}
\newblock \bibinfo{title}{Current-driven domain wall motion along high
  perpendicular anisotropy multilayers: The role of the rashba field, the spin
  hall effect, and the dzyaloshinskii-moriya interaction}.
\newblock \emph{\bibinfo{journal}{Applied Physics Letters}}
  \textbf{\bibinfo{volume}{103}}, \bibinfo{pages}{072406}
  (\bibinfo{year}{2013}).

\bibitem{Fan2014}
\bibinfo{author}{Fan, X.} \emph{et~al.}
\newblock \bibinfo{title}{Quantifying interface and bulk contributions to
  spin-orbit torque in magnetic bilayers}.
\newblock \emph{\bibinfo{journal}{Nature Communications}}
  \textbf{\bibinfo{volume}{5}}, \bibinfo{pages}{3042} (\bibinfo{year}{2014}).

\bibitem{HinSim2014}
\bibinfo{author}{Hin~Sim, C.}, \bibinfo{author}{Cheng~Huang, J.},
  \bibinfo{author}{Tran, M.} \& \bibinfo{author}{Eason, K.}
\newblock \bibinfo{title}{Asymmetry in effective fields of spin-orbit torques
  in pt/co/pt stacks}.
\newblock \emph{\bibinfo{journal}{Applied Physics Letters}}
  \textbf{\bibinfo{volume}{104}}, \bibinfo{pages}{--} (\bibinfo{year}{2014}).

\bibitem{Garello2013}
\bibinfo{author}{Garello, K.} \emph{et~al.}
\newblock \bibinfo{title}{Symmetry and magnitude of spin-orbit torques in
  ferromagnetic heterostructures}.
\newblock \emph{\bibinfo{journal}{Nat Nano}} \textbf{\bibinfo{volume}{8}},
  \bibinfo{pages}{587--593} (\bibinfo{year}{2013}).

\bibitem{Zhang2015}
\bibinfo{author}{Zhang, W.}, \bibinfo{author}{Han, W.}, \bibinfo{author}{Jiang,
  X.}, \bibinfo{author}{Yang, S.-H.} \& \bibinfo{author}{S.~P.~Parkin, S.}
\newblock \bibinfo{title}{Role of transparency of platinum-ferromagnet
  interfaces in determining the intrinsic magnitude of the spin hall effect}.
\newblock \emph{\bibinfo{journal}{Nat Phys}} \textbf{\bibinfo{volume}{11}},
  \bibinfo{pages}{496--502} (\bibinfo{year}{2015}).

\bibitem{Tokac2015}
\bibinfo{author}{Tokac, M.} \emph{et~al.}
\newblock \bibinfo{title}{Interfacial structure dependent spin mixing
  conductance in cobalt thin films}.
\newblock \emph{\bibinfo{journal}{Phys. Rev. Lett.}}
  \textbf{\bibinfo{volume}{115}}, \bibinfo{pages}{056601--}
  (\bibinfo{year}{2015}).

\bibitem{Ryu2013}
\bibinfo{author}{Ryu, K.-S.}, \bibinfo{author}{Thomas, L.},
  \bibinfo{author}{Yang, S.-H.} \& \bibinfo{author}{Parkin, S.}
\newblock \bibinfo{title}{Chiral spin torque at magnetic domain walls}.
\newblock \emph{\bibinfo{journal}{Nat Nano}} \textbf{\bibinfo{volume}{8}},
  \bibinfo{pages}{527--533} (\bibinfo{year}{2013}).

\bibitem{Ryu2014}
\bibinfo{author}{Ryu, K.-S.}, \bibinfo{author}{Yang, S.-H.},
  \bibinfo{author}{Thomas, L.} \& \bibinfo{author}{Parkin, S. S.~P.}
\newblock \bibinfo{title}{Chiral spin torque arising from proximity-induced
  magnetization}.
\newblock \emph{\bibinfo{journal}{Nat Commun}} \textbf{\bibinfo{volume}{5}},
  \bibinfo{pages}{--} (\bibinfo{year}{2014}).

\bibitem{Yang2015}
\bibinfo{author}{Yang, H.}, \bibinfo{author}{Thiaville, A.},
  \bibinfo{author}{Rohart, S.}, \bibinfo{author}{Fert, A.} \&
  \bibinfo{author}{Chshiev, M.}
\newblock \bibinfo{title}{Anatomy of dzyaloshinskii-moriya interaction at
  $\mathrm{Co}/\mathrm{Pt}$ interfaces}.
\newblock \emph{\bibinfo{journal}{Phys. Rev. Lett.}}
  \textbf{\bibinfo{volume}{115}}, \bibinfo{pages}{267210}
  (\bibinfo{year}{2015}).

\bibitem{Wienke1991}
\bibinfo{author}{Wienke, R.}, \bibinfo{author}{Sch\"{u}tz, G.} \&
  \bibinfo{author}{Ebert, H.}
\newblock \bibinfo{title}{Determination of local magnetic moments of 5d
  impurities in fe detected via spin dependent absorption}.
\newblock \emph{\bibinfo{journal}{J. Appl. Phys.}}
  \textbf{\bibinfo{volume}{69}}, \bibinfo{pages}{6147--6149}
  (\bibinfo{year}{1991}).

\bibitem{burn2014focused}
\bibinfo{author}{Burn, D.}, \bibinfo{author}{Hase, T.} \&
  \bibinfo{author}{Atkinson, D.}
\newblock \bibinfo{title}{Focused-ion-beam induced interfacial intermixing of
  magnetic bilayers for nanoscale control of magnetic properties}.
\newblock \emph{\bibinfo{journal}{Journal of Physics: Condensed Matter}}
  \textbf{\bibinfo{volume}{26}}, \bibinfo{pages}{236002}
  (\bibinfo{year}{2014}).

\bibitem{Kim2016}
\bibinfo{author}{Kim, N.~H.} \emph{et~al.}
\newblock \bibinfo{title}{interfacial dzyaloshinskii-moriya interaction,
  surface anisotropy energy, and spin pumping at spin orbit coupled ir/co
  interface}.
\newblock \emph{\bibinfo{journal}{Applied Physics Letters}}
  \textbf{\bibinfo{volume}{108}} (\bibinfo{year}{2016}).
\newblock \urlprefix\url{http://dx.doi.org/10.1063/1.4945685}.

\bibitem{Stashkevich2015}
\bibinfo{author}{Stashkevich, A.~A.} \emph{et~al.}
\newblock \bibinfo{title}{Experimental study of spin-wave dispersion in py/pt
  film structures in the presence of an interface dzyaloshinskii-moriya
  interaction}.
\newblock \emph{\bibinfo{journal}{Phys. Rev. B}} \textbf{\bibinfo{volume}{91}},
  \bibinfo{pages}{214409--} (\bibinfo{year}{2015}).

\bibitem{Udvardi2009}
\bibinfo{author}{Udvardi, L.} \& \bibinfo{author}{Szunyogh, L.}
\newblock \bibinfo{title}{Chiral asymmetry of the spin-wave spectra in
  ultrathin magnetic films}.
\newblock \emph{\bibinfo{journal}{Phys. Rev. Lett.}}
  \textbf{\bibinfo{volume}{102}}, \bibinfo{pages}{207204}
  (\bibinfo{year}{2009}).

\bibitem{Cortes2013}
\bibinfo{author}{Cort\'{e}s-Ortu\~{n}o, D.} \& \bibinfo{author}{Landeros, P.}
\newblock \bibinfo{title}{Influence of the dzyaloshinskii-moriya interaction on
  the spin-wave spectra of thin films}.
\newblock \emph{\bibinfo{journal}{Journal of Physics: Condensed Matter}}
  \textbf{\bibinfo{volume}{25}}, \bibinfo{pages}{156001}
  (\bibinfo{year}{2013}).

\bibitem{Moon2013}
\bibinfo{author}{Moon, J.-H.} \emph{et~al.}
\newblock \bibinfo{title}{Spin-wave propagation in the presence of interfacial
  dzyaloshinskii-moriya interaction}.
\newblock \emph{\bibinfo{journal}{Phys. Rev. B}} \textbf{\bibinfo{volume}{88}},
  \bibinfo{pages}{184404} (\bibinfo{year}{2013}).

\bibitem{Kost2014}
\bibinfo{author}{Kostylev, M.}
\newblock \bibinfo{title}{Interface boundary conditions for dynamic
  magnetization and spin wave dynamics in a ferromagnetic layer with the
  interface dzyaloshinskii-moriya interaction}.
\newblock \emph{\bibinfo{journal}{J. Appl. Phys.}}
  \textbf{\bibinfo{volume}{115}}, \bibinfo{pages}{233902}
  (\bibinfo{year}{2014}).

\bibitem{Zakeri2010}
\bibinfo{author}{Zakeri, K.} \emph{et~al.}
\newblock \bibinfo{title}{Asymmetric spin-wave dispersion on fe(110): Direct
  evidence of the dzyaloshinskii-moriya interaction}.
\newblock \emph{\bibinfo{journal}{Phys. Rev. Lett.}}
  \textbf{\bibinfo{volume}{104}}, \bibinfo{pages}{137203}
  (\bibinfo{year}{2010}).

\bibitem{Di2015}
\bibinfo{author}{Di, K.} \emph{et~al.}
\newblock \bibinfo{title}{Direct observation of the dzyaloshinskii-moriya
  interaction in a pt/co/ni film}.
\newblock \emph{\bibinfo{journal}{Phys. Rev. Lett.}}
  \textbf{\bibinfo{volume}{114}}, \bibinfo{pages}{047201}
  (\bibinfo{year}{2015}).

\bibitem{Belmeguenai2015}
\bibinfo{author}{Belmeguenai, M.} \emph{et~al.}
\newblock \bibinfo{title}{Interfacial dzyaloshinskii-moriya interaction in
  perpendicularly magnetized ${\text{pt/co/alo}}_{x}$ ultrathin films measured
  by brillouin light spectroscopy}.
\newblock \emph{\bibinfo{journal}{Phys. Rev. B}} \textbf{\bibinfo{volume}{91}},
  \bibinfo{pages}{180405} (\bibinfo{year}{2015}).

\bibitem{Cho2015}
\bibinfo{author}{Cho, J.} \emph{et~al.}
\newblock \bibinfo{title}{Thickness dependence of the interfacial
  dzyaloshinskii-moriya interaction in inversion symmetry broken systems}.
\newblock \emph{\bibinfo{journal}{Nat. Commun.}} \textbf{\bibinfo{volume}{6}},
  \bibinfo{pages}{--} (\bibinfo{year}{2015}).

\bibitem{Gurevich1996}
\bibinfo{author}{Gurevich, A.~G.} \& \bibinfo{author}{Melkov, G.~A.}
\newblock \emph{\bibinfo{title}{Magnetization Oscillations and Waves}}
  (\bibinfo{publisher}{CRC Press, New York}, \bibinfo{year}{1996}).

\bibitem{Hrabec2014}
\bibinfo{author}{Hrabec, A.} \emph{et~al.}
\newblock \bibinfo{title}{Measuring and tailoring the dzyaloshinskii-moriya
  interaction in perpendicularly magnetized thin films}.
\newblock \emph{\bibinfo{journal}{Phys. Rev. B}} \textbf{\bibinfo{volume}{90}},
  \bibinfo{pages}{020402--} (\bibinfo{year}{2014}).

\bibitem{Belmeguenai2016}
\bibinfo{author}{Belmeguenai, M.} \emph{et~al.}
\newblock \bibinfo{title}{Brillouin light scattering investigation of the
  thickness dependence of dzyaloshinskii-moriya interaction in
  $\mathrm{C}{\mathrm{o}}_{0.5}\mathrm{F}{\mathrm{e}}_{0.5}$ ultrathin films}.
\newblock \emph{\bibinfo{journal}{Phys. Rev. B}} \textbf{\bibinfo{volume}{93}},
  \bibinfo{pages}{174407} (\bibinfo{year}{2016}).

\bibitem{Azzawi2016}
\bibinfo{author}{Azzawi, S.} \emph{et~al.}
\newblock \bibinfo{title}{Evolution of damping in ferromagnetic/nonmagnetic
  thin film bilayers as a function of nonmagnetic layer thickness}.
\newblock \emph{\bibinfo{journal}{Phys. Rev. B}} \textbf{\bibinfo{volume}{93}},
  \bibinfo{pages}{054402} (\bibinfo{year}{2016}).

\bibitem{denBroeder1991}
\bibinfo{author}{den Broeder, F.}, \bibinfo{author}{Hoving, W.} \&
  \bibinfo{author}{Bloemen, P.}
\newblock \bibinfo{title}{Magnetic anisotropy of multilayers}.
\newblock \emph{\bibinfo{journal}{J. Magn. Magn. Mater.}}
  \textbf{\bibinfo{volume}{93}}, \bibinfo{pages}{562--570}
  (\bibinfo{year}{1991}).

\bibitem{Soucaille2016}
\bibinfo{author}{Soucaille, R.} \emph{et~al.}
\newblock \emph{\bibinfo{journal}{Phys Rev B}} \textbf{\bibinfo{volume}{94}},
  \bibinfo{pages}{104431} (\bibinfo{year}{2016}).

\bibitem{Macke2014}
\bibinfo{author}{Macke, S.} \& \bibinfo{author}{Goering, E.}
\newblock \bibinfo{title}{Magnetic reflectometry of heterostructures}.
\newblock \emph{\bibinfo{journal}{Journal of Physics-condensed Matter}}
  \textbf{\bibinfo{volume}{26}}, \bibinfo{pages}{363201}
  (\bibinfo{year}{2014}).

\bibitem{Stepanov2000}
\bibinfo{author}{Stepanov, S.~A.} \& \bibinfo{author}{Sinha, S.~K.}
\newblock \bibinfo{title}{X-ray resonant reflection from magnetic multilayers:
  Recursion matrix algorithm}.
\newblock \emph{\bibinfo{journal}{Phys. Rev. B}} \textbf{\bibinfo{volume}{61}},
  \bibinfo{pages}{15302--15311} (\bibinfo{year}{2000}).

\bibitem{Kim2016a}
\bibinfo{author}{Kim, D.-O.} \emph{et~al.}
\newblock \bibinfo{title}{Asymmetric magnetic proximity effect in a pd/co/pd
  trilayer system}.
\newblock \emph{\bibinfo{journal}{Scientific Reports}}
  \textbf{\bibinfo{volume}{6}}, \bibinfo{pages}{25391} (\bibinfo{year}{2016}).
\newblock \urlprefix\url{http://dx.doi.org/10.1038/srep25391}.

\bibitem{Larde2009}
\bibinfo{author}{Lard\'{e}, R.} \emph{et~al.}
\newblock \bibinfo{title}{Structural analysis of a (pt/co)3/irmn multilayer:
  Investigation of sub-nanometric layers by tomographic atom probe}.
\newblock \emph{\bibinfo{journal}{J. Appl. Phys.}}
  \textbf{\bibinfo{volume}{105}}, \bibinfo{pages}{69--73}
  (\bibinfo{year}{2009}).

\bibitem{Okamoto1985}
\bibinfo{author}{Okamoto, H.} \& \bibinfo{author}{Massalski, T.~B.}
\newblock \bibinfo{title}{The au-pt (gold-platinum) system}.
\newblock \emph{\bibinfo{journal}{Bulletin of Alloy Phase Diagrams}}
  \textbf{\bibinfo{volume}{6}}, \bibinfo{pages}{46--56} (\bibinfo{year}{1985}).

\bibitem{Bharadwaj1995}
\bibinfo{author}{Bharadwaj, S.~R.}, \bibinfo{author}{Tripathi, S.~N.} \&
  \bibinfo{author}{Chandrasekharaiah, M.~S.}
\newblock \bibinfo{title}{The ir-pt (iridium-platinum) system}.
\newblock \emph{\bibinfo{journal}{Journal of Phase Equilibria}}
  \textbf{\bibinfo{volume}{16}}, \bibinfo{pages}{460--464}
  (\bibinfo{year}{1995}).

\bibitem{Fert1980}
\bibinfo{author}{Fert, A.} \& \bibinfo{author}{Levy, P.-M.}
\newblock \emph{\bibinfo{journal}{Phys Rev Lett}}
  \textbf{\bibinfo{volume}{44}}, \bibinfo{pages}{1538} (\bibinfo{year}{1980}).

\bibitem{Balten1990}
\bibinfo{author}{Baltensberger, W.} \& \bibinfo{author}{Helman, J.-S.}
\newblock \emph{\bibinfo{journal}{Appl Phys Lett}}
  \textbf{\bibinfo{volume}{57}}, \bibinfo{pages}{2954} (\bibinfo{year}{1990}).

\bibitem{Purcell1992}
\bibinfo{author}{Purcell, S.-T.}, \bibinfo{author}{Johnson, M.-T.},
  \bibinfo{author}{McGee, N.-W.-E.}, \bibinfo{author}{Cochoorn, R.} \&
  \bibinfo{author}{Hoving, W.}
\newblock \emph{\bibinfo{journal}{Phys Rev B}} \textbf{\bibinfo{volume}{45}},
  \bibinfo{pages}{13064} (\bibinfo{year}{1992}).

\bibitem{Johnson1992}
\bibinfo{author}{Johnson, M.-T.} \emph{et~al.}
\newblock \emph{\bibinfo{journal}{Phys Rev Lett}}
  \textbf{\bibinfo{volume}{68}}, \bibinfo{pages}{2688} (\bibinfo{year}{1992}).

\bibitem{Fuss1992}
\bibinfo{author}{Fuss, A.}, \bibinfo{author}{Demokritov, S.},
  \bibinfo{author}{Gr\"unberg, P.} \& \bibinfo{author}{Zinn, W.}
\newblock \emph{\bibinfo{journal}{J Magn Magn Mater}}
  \textbf{\bibinfo{volume}{103}}, \bibinfo{pages}{L221} (\bibinfo{year}{1992}).

\bibitem{Bruno1992}
\bibinfo{author}{Bruno, P.} \& \bibinfo{author}{Chappert, C.}
\newblock \emph{\bibinfo{journal}{Phys Rev B}} \textbf{\bibinfo{volume}{46}},
  \bibinfo{pages}{261} (\bibinfo{year}{1992}).

\bibitem{Bjorck2014}
\bibinfo{author}{Bj{\"{o}}rck, M.} \emph{et~al.}
\newblock \bibinfo{title}{Reflectivity studies of magnetic heterostructures}.
\newblock \emph{\bibinfo{journal}{Journal of Surfaces and Interfaces of
  Materials}} \textbf{\bibinfo{volume}{2}}, \bibinfo{pages}{24--32}
  (\bibinfo{year}{2014}).

\bibitem{Bjorck2011}
\bibinfo{author}{Bj{\"{o}}rck, M.}
\newblock \bibinfo{title}{Fitting with differential evolution: an introduction
  and evaluation}.
\newblock \emph{\bibinfo{journal}{Journal of Applied Crystallography}}
  \textbf{\bibinfo{volume}{44}}, \bibinfo{pages}{1198--1204}
  (\bibinfo{year}{2011}).

\bibitem{CamleyMills}
\bibinfo{author}{Camley, R.~E.} \& \bibinfo{author}{Mills, D.~L.}
\newblock \emph{\bibinfo{journal}{Phys. Rev. B}} \textbf{\bibinfo{volume}{18}},
  \bibinfo{pages}{4821} (\bibinfo{year}{1978}).

\end{thebibliography}

\section*{Acknowledgements}
Funding is acknowledged from EPSRC for the studentship for R.M.~Rowan-Robinson (1212684) and Grant Ref.~EP/L000121/1. We acknowledge the funding for beam time awarded on the EPSRC-funded XMaS Beamline at the ESRF for the PIM measurements and we are grateful for the contributions of Dr Laurence Bouchenoire at XMaS and Dr Matts Bj\"{o}rk for insights and support with GenX analysis of the x-ray reflectivity.
Data supporting this work is available at http://dx.doi.org/10.15128/r1bv73c0401.

\section*{Author contributions statement}

DA and AS together conceived the study. RMRR prepared the samples with input from ATH and DA. AS, YR, MB and MC carried out the BLS experiments. YR and AS processed the raw BLS data. RMRR, ATH, TPAH and DA undertook the PIM measurements. RMRR undertook the PIM analysis with input from TPAH and DA. RMRR, AS, AT and DA wrote the manuscript. All authors discussed the results and analysis and commented on the manuscript.


\section*{Additional information}

The corresponding author is responsible for submitting a \href{http://www.nature.com/srep/policies/index.html#competing}{competing financial interests statement} on behalf of all authors of the paper. This statement must be included in the submitted article file.

\newpage


\begin{figure}
\centering
\includegraphics[width=0.6\textwidth]{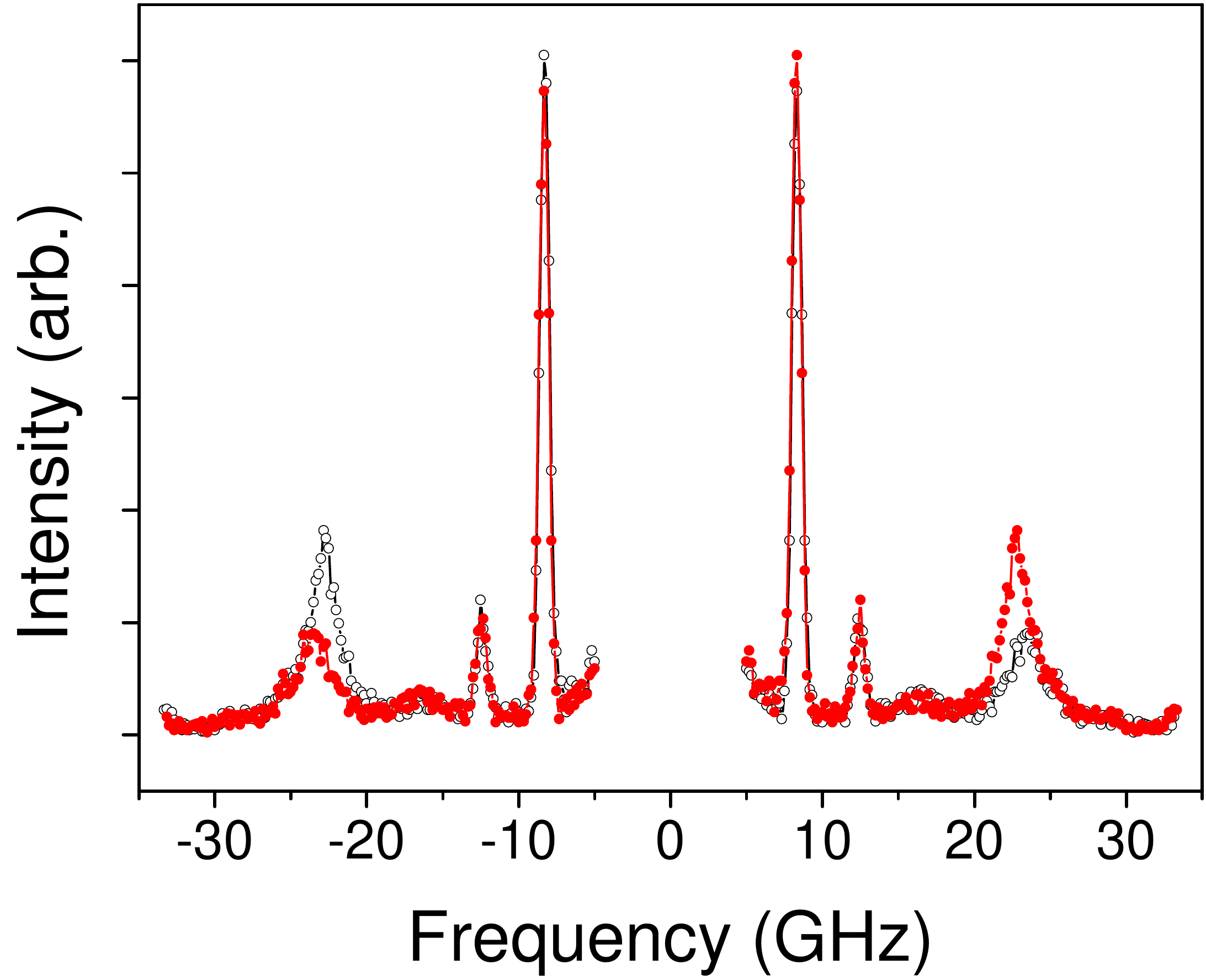}
\caption{Measured BLS spectrum obtained in the absence of the analyser. Data shown by black (open) markers are obtained at +4~kOe in-plane field and at $\theta= 50 \deg$. Data in red (filled) markers show the +4~kOe field spectrum reflected.}
\label{figmethod}
\end{figure}

\end{document}